# Coherent Control of Rydberg States in Silicon,

P. T. Greenland1, S. A. Lynch1, A. F. G. van der Meer2, , B. N. Murdin3, C. R. Pidgeon4 B. Redlich2 N. Q. Vinh2,† , G. Aeppli1

*1London Centre for Nanotechnology and Department of Physics and Astronomy, University College London, WC1H 0AH, England*

*2FOM Institute for Plasma Physics "Rijnhuizen", P.O. Box 1207, NL-3430 BE Nieuwegein, The Netherlands*

*3Advanced Technology Institute, University of Surrey, Guildford GU2 7XH, England*

*4Heriot-Watt University, Department of Physics, Riccarton, Edinburgh, EH14 4AS, Scotland*

**Laser cooling and electromagnetic traps have led to a revolution in atomic physics, yielding dramatic discoveries from Bose-Einstein condensation to quantum control of single atoms[1]. Of particular interest because they can be exploited for quantum control of one atom by another are excited Rydberg states[2,3,4], where wavefunctions are expanded from their ground state extents of below 0.1 nm to several nm and even beyond; this allows atoms far enough apart to be non-interacting in their ground states to strongly interact in their excited states. For eventual application[5], a solid state implementation is very desirable, and we demonstrate here the coherent control of impurity wavefunctions in the most ubiquitous donor in a semiconductor, namely phosphorous-doped silicon. Our experiments take advantage of a free electron laser to stimulate and observe photon echoes[6,7], the orbital analog of the Hahn spin echo[8], and Rabi oscillations**

† Present address ITST, Department of Physics, University of California, Santa Barbara CA 93106-4170

**familiar from magnetic resonance spectroscopy. As well as extending atomic physicists' explorations[1,2,3,9] of quantum phenomena to the solid sate, our work adds coherent THz radiation as a particularly precise regulator of orbitals in solids to the list of controls, such as pressure and chemical composition, already familiar to materials scientists[10].**

There are several approaches to atom trap physics in solids. One possibility is to use quantum dots hosted by compound semiconductors[11,12], which have the advantage of addressability using conventional lasers. However, this relies on expensive fabrication and experiments have been mostly restricted to the frequency domain. We follow here another avenue, which requires more exotic laser technology but lends itself to straightforward time-domain measurements and has much simpler sample requirements. We use the ubiquitous semiconductor-donor combination Si:P , which thus opens up the possibility of exploiting sophisticated semiconductor device processing technologies for combined electrical and coherent optical control.

Our experiments exploit the remarkable fact that impurities with one more valence electron than the host semiconductor display ground and excited states corresponding to the free hydrogen atom Rydberg series, the touchstone for the Bohr formulation of quantum mechanics. In the semiconductor[13], the conduction bands, with curvature characterized by the electron effective mass $m^*$, play the role of the free electron continuum, while the Coulomb interaction is reduced in proportion to its dielectric constant $\varepsilon_r$ . Thus, the binding energy scales with $m^*/\varepsilon_r^2$ while the orbital radii scale with $\varepsilon_r/m^*$. For typical donors in Si, the lowest energy Lyman series line is therefore in the THz regime – in the case of phosphorus (P) the $1sA \rightarrow 2p_0$ transition is at 34.2meV, equivalent to 36.2μm and 8.29THz (Fig 1a). There are also smaller level splittings associated with the broken rotational symmetry in the solid. The orbitals have an order of magnitude larger spatial extent than those for hydrogen in vacuum: the $2p_0$ level, for example has an extent of ~ 10 nm, enclosing about $10^4$ Si atoms, and is thus comparable in size to transistors already in commercial use. Previous frequency[14] and time-domain studies[15] have established the astonishing longevity of the excited states, with a population lifetime $T_1$ of 200 ps and corresponding oscillator quality factors of 2000 or more.

A two-level atom resonantly illuminated by the high intensity coherent light from a laser undergoes Rabi oscillations at a frequency given by $\Omega = F_0\mu_{12}/\hbar$ where $F_0$ is the electric field envelope of the light beam and $\mu_{12}$ is the transition dipole matrix element. For a pulse of finite duration, the excited state polarization that remains in the system after the pulse has passed varies sinusoidally with the pulse area, $A=\mu_{12}/\hbar \int F(t)dt$. If the laser is at resonance with the 1s to $2p_0$ transition it will produce a linear superposition of 1s and $2p_0$ wavefunctions – a very simple wavepacket which oscillates in time as the superposition precesses around the Bloch sphere (see fig 1b), representing the quantum mechanical state space for two-level systems. For an ensemble, all the wavepackets initially radiate in phase, and therefore strongly, to produce coherent radiation. This coherence is lost, owing to small offsets in the resonant frequencies resulting from differences in the local environment, and the radiation weakens as the dipoles dephase on a timescale given by the inverse of the 1s-$2p_0$ inhomogeneous linewidth, measureable in the frequency domain using conventional continuous wave infrared spectroscopy. However, their relative phases can be restored by a subsequent laser pulse leading to a second burst of coherent radiation – the photon echo – which appears later by a time equal to the time difference between the initial and rephasing pulses, in precise analogy to the well-known Hahn spin echo[7] (fig 1c). In general, the amplitude of the echo will decrease as function of time delay of the rephasing pulse with a characteristic time $T_2$, due to population decay and stochastic phase jumps of the oscillators. The emission appears not only at a well-defined time, but also at a well-defined angle: its direction is given by the vector equation $\boldsymbol{k}_E = 2\boldsymbol{k}_2 - \boldsymbol{k}_1$ where $\boldsymbol{k}_1$ is the wave-vector of the pump, $\boldsymbol{k}_2$ is that of the rephasing beam[6]. This means that a genuine echo, as opposed to more conventional four-wave mixing effects where different coherent beams are present in the sample simultaneously, has signatures in space and time. For our measurements to discover the orbital echo in Si:P, we accordingly set out to establish both the direction of the echo beam relative to the pump and rephasing beams (Fig 2a ANGLE) and also the arrival time of the echo relative to the pump and rephasing pulses by interfering each of them with a reference pulse (Fig 2a TIMING).

We performed most of our experiments on a Czochralski-grown 110 natural Si wafer, of thickness 200μ and doped with $1.5 \times 10^{15}$ P donors/cm$^3$, but also verified the key results on several other samples, as described in the Supplementary Material. The THz source was the FELIX free electron laser at the FOM institute in Nieuwegein[16], which produces trains of radiation pulses that for the present experiment were tuned so that their frequency matched the $1s_A \rightarrow 2p_0$ transition of Si:P and had durations of ~10 ps, as also verified below.

The first experiment was simply to measure the angular distribution of intensity emitted by the sample. We did this by recording the intensity while moving a small mirror across the collecting parabolic mirror. Fig. 2b shows the resulting profile, which does indeed show the expected echo peak at $\theta_{echo}$. There are also sharp maxima at angles corresponding to the directions of pump and rephasing pulses.

Having shown the appearance of an echo signal with the correct wavevector, we turn to verification of its arrival time, which we determine – taking advantage of the coherence of the radiation from the free electron laser – using a reference pulse split from the rephasing pulse and a delay line. The transmitted pump, rephasing and emitted echo pulses, as well as the reference pulse are all focussed onto the detector through a pinhole to produce a characteristic interference pattern in time. We exploit the angular dispersion of the pump, rephasing and echo pulses and block all but one of them, thereby obtaining the interference patterns of the reference beam with the pump, rephasing and echo beams separately. By subtracting the mean intensity and squaring the result, the arrival times and shapes of the pump, rephasing and echo pulses can then be determined as a function of time (Fig. 2c). All three pulses take the form of well-defined peaks, with the maxima occurring at the times (Fig. 2d) anticipated for echos.

We can now measure the dephasing time, $T_2$, by observing the dependence of the echo intensity on the time difference $\tau_{12}$ between the pump and rephasing pulses[17]. As shown in Fig. 2e, the intensity of the echo decays exponentially as exp $-\tau_{12}/T_{exp}$ and $T_2=4T_{exp}$, where the factor of 4 arises because the time between the emission of the echo pulse and the pump pulse that caused it is twice the pump-rephasing pulse delay, and the intensity of the echo decays twice as fast as the polarization amplitude.

The value of $T_2$ at low laser intensity was 160 ±20 ps. The laser excitation introduces additional sources of decoherence, and so we expect that $T_2$ should fall as the laser intensity increases. Fig 3a demonstrates this – a particular echo decay is shown in the inset. The extra decoherence arises because electrons produced by multiphoton excitation of some donors can collide with the un-ionized oscillators which produce the echo, and so increase their dephasing rate. The effects of this were calculated (see Supplementary Materials) using a two-level reduced density matrix[18], including photoionization and photoelectron collisions, which represents a slight extension of the more usual Bloch equations for the spin echo. The spatial profile of the laser beam must also be considered because it will induce a corresponding spatial distribution of coherent excited state polarizations as well as incoherent processes. The density matrix calculations predict the behaviour of $T_{exp}$ shown in Fig. 3a, and reproduce the experimental observations well.

We have demonstrated orbital echos and long decoherence times for Si:P. To determine how well we can actually control impurity wavefunctions, i.e. the extent to which we can dial in coherent superpositions of different Rydberg states, we tracked Rabi oscillations, in the standard echo detection mode, by measuring the magnitude of the echo as a function of pump pulse area. Results for several pulse durations and rephasing pulse areas were obtained (see supplementary material). Fig 3b shows that the experimental echo intensity (magenta squares in the figure) actually displays one complete oscillation, and agrees well with the theoretical prediction (magenta line), which takes into account the decoherence mechanisms also needed to account for $T_2$, as above, and is indicated by the dashed line in fig. 3b. Fig. 3b also shows what would be observed if the extra decoherence, and spatial intensity variation were absent.

Using the theory to fit many results of this type, and taking into account the beam attenuation due to transmission through the cryostat window, and the Si air interface gives a dipole matrix element, $\mu_{12}$=0.28 ± 0.03 nm. This is much lower than the scaled hydrogenic value, a consequence of the central cell correction[19], but comparable to values derived from low field absorption measurements[20], which give values in the 0.33 - 0.5 nm range. Since our value relies on an absolute measurement of the FELIX pulse energy – notoriously unreliable at THz frequencies – the agreement is satisfactory. For the photoionization cross-section from the excited state we find $\sigma=1.28\times10^{-20}$ m$^2$, about twice what would be expected from hydrogenic scaling, and for the collision cross-section for free electrons with un-ionised donors we obtain $\sigma_e= 8\times10^{-16}$ m$^2$. This is similar to electron-donor recombination cross sections in Si[21].

We have directly observed photon echoes and Rabi oscillations produced by coherent optical excitation of phosphorus donors in silicon with intense THz pulses from a free-electron laser. Fig 4, which compares Si:P to an isolated H atom, summarizes the key parameters that we have deduced from our experiments. All electromagnetic parameters scale to within factors of two to what is expected based on the Bohr model with renormalized effective mass and dielectric constant, except the 1s→2p dipole matrix element, which is affected by central cell corrections. This includes not only the Rydberg series itself but also the photoionization cross-section of the $2p_0$ state. The only substantive differences between free hydrogen and Si:P are then the much shorter $T_1$ and $T_2$ times for the latter, due to phonons which are characteristic of a solid, but not of the vacuum.

Our work shows that we can prepare coherent mixtures of different orbital states for one of the most common impurities in the most common semiconductor. These mixtures have dephasing times $T_2$ in excess of 100 ps, three orders of magnitude larger than the 100 femtoseconds corresponding to the frequency of the transition between the orbital states. The frequency domain linewidth associated with our measured $T_2$ of 160 ps is 8.2 μeV. This is about twice the linewidth reported for P in isotopically pure float zone Si[14], so there is reason to believe that more carefully prepared samples will have a longer $T_2$ than the Czochralski Si used here.

Coherent control of donor orbitals in silicon opens up many possibilities currently under examination using atom traps[1,2,3], such as entanglement of pairs of impurities whose ground state wavefunctions are too compact to interact. Modern nanotechnology, which has recently been used for deterministic positioning of individual impurities in silicon[22], will also enable such control to be used for the regulation of magnetism[5] by opening and closing exchange pathways by the timed preparation of excited states.

**Supplementary Information** accompanies the paper on **www.nature.com/nature**

**Acknowledgements**

We are grateful for helpful conversations with A.J. Fisher, A.M. Stoneham, C. Kay and G. Morley, R. Hulet for pointing out reference [8], and experimental assistance from K Litvinenko and G. Morley, and the financial support of NWO and EPSRC.

**Author Statement**

NQV and CRP initiated this work, PTG, SAL, BNM, NQV, and GA designed the research programme; NQV, PTG, SAL, LvdM and BR performed the experiments; PTG performed the theory and analysis; PTG, BNM, SAL and GA wrote the paper.

Correspondence should be addressed to PTG (ptg@globalnet.co.uk).

---

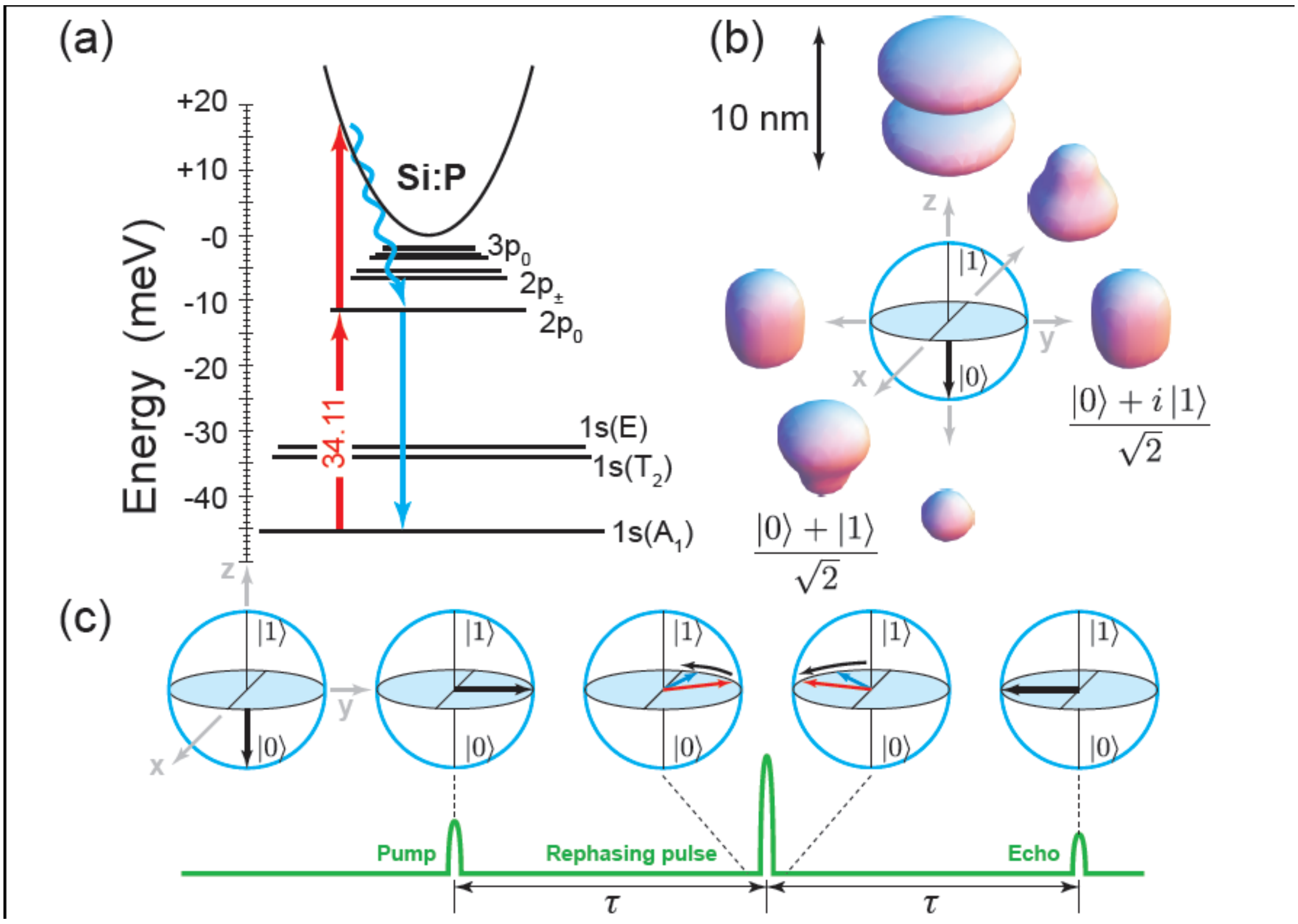

Fig 1(a) The spectrum of an isolated P donor in Si. The primary excitation and two-photon ionization paths are shown in red; dephasing paths – photoelectron collisions and phonon decay – are shown in blue. (b) The Bloch sphere, with some sample wavepackets. The ground 1sA state is at the south pole, the $2p_0$ excited state at the north pole. Round the equator we show how the wavepacket varies as the relative phase of a 50-50 mixture evolves in time. This is the time dependent combination we excite to make the photon echo. (c) The classic Hahn sequence, and the corresponding behaviour of the Bloch vector. Ideally the first pulse has an area of π/2, and the second an area of π.

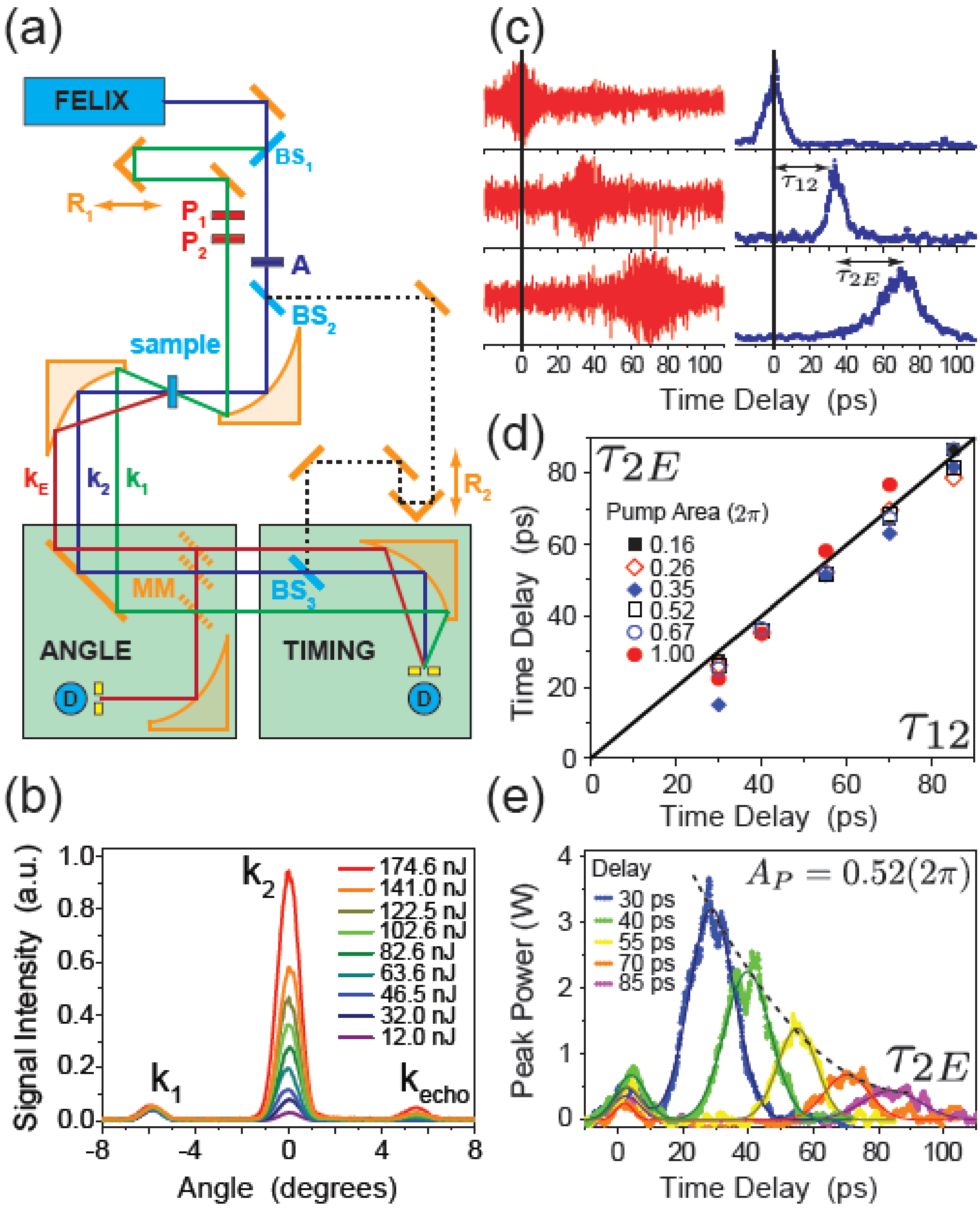

Fig 2 (a) Schematic showing the experimental geometry. The pump ($\boldsymbol{k}_1$) beam (green) and rephasing ($\boldsymbol{k}_2$) beam (blue) intersect in the sample (at ~5° in the experiment) and the echo $\boldsymbol{k}_E$ (orange) is emitted in the direction 2$\boldsymbol{k}_2$-$\boldsymbol{k}_1$. Although the paths are shown correctly, their total lengths are not drawn to scale. (b) The intensities of the spatially dispersed signals are recorded by translating the detector across the far-field. Simple geometry enables the spatial dispersion to be converted to an angle of emission to show that $\boldsymbol{k}_E = 2\boldsymbol{k}_2 - \boldsymbol{k}_1$ as predicted (c) Result of cross-correlation of a reference beam (***k****) (dashed) with the pump $\boldsymbol{k_1}$ (top), the rephasing pulse $\boldsymbol{k}_2$ (centre) and the echo $\boldsymbol{k_E}$ (bottom) by interfering them on the detector. The abscissa is the arrival time of the reference pulse ***k**** in ps, relative to an arbitrary zero. On the left is the detector signal showing the interference pattern. A moving average has been subtracted, in order to remove the background and laser drift. The pump, rephasing and echo temporal profiles can be obtained from the square of these interference patterns. Gaussian fits of the pump and rephasing pulses had full-widths-at-half-maximum intensity of 7.7±1.5ps, consistent with the inverse of the spectral width of 0.28%, showing that the pulses are indeed band-width limited. The echo duration, 27.8±7.6ps, is somewhat longer than would be expected from the measured inhomogeneous frequency domain line-width (~ 200 μeV). (d) The echo arrival time $\tau_{2E}$ as a function of $\tau_{12}$ derived in the same way for several different pump areas, showing that, within experimental error, the echo arrives when expected. Finally (e) shows a set of echo profiles for different pump-rephasing pulse delays $\tau_{12}$. The ordinate is a very crude estimate of the echo power in W; the temporal width of the echo implies an inhomogeneous linewidth of ~400 μeV, suggesting the sample is strained. The black line is an exponential fit, with time constant $T_{exp}$ =28.4 ps showing that the echo intensity falls with $\tau_{12,}$ as expected.

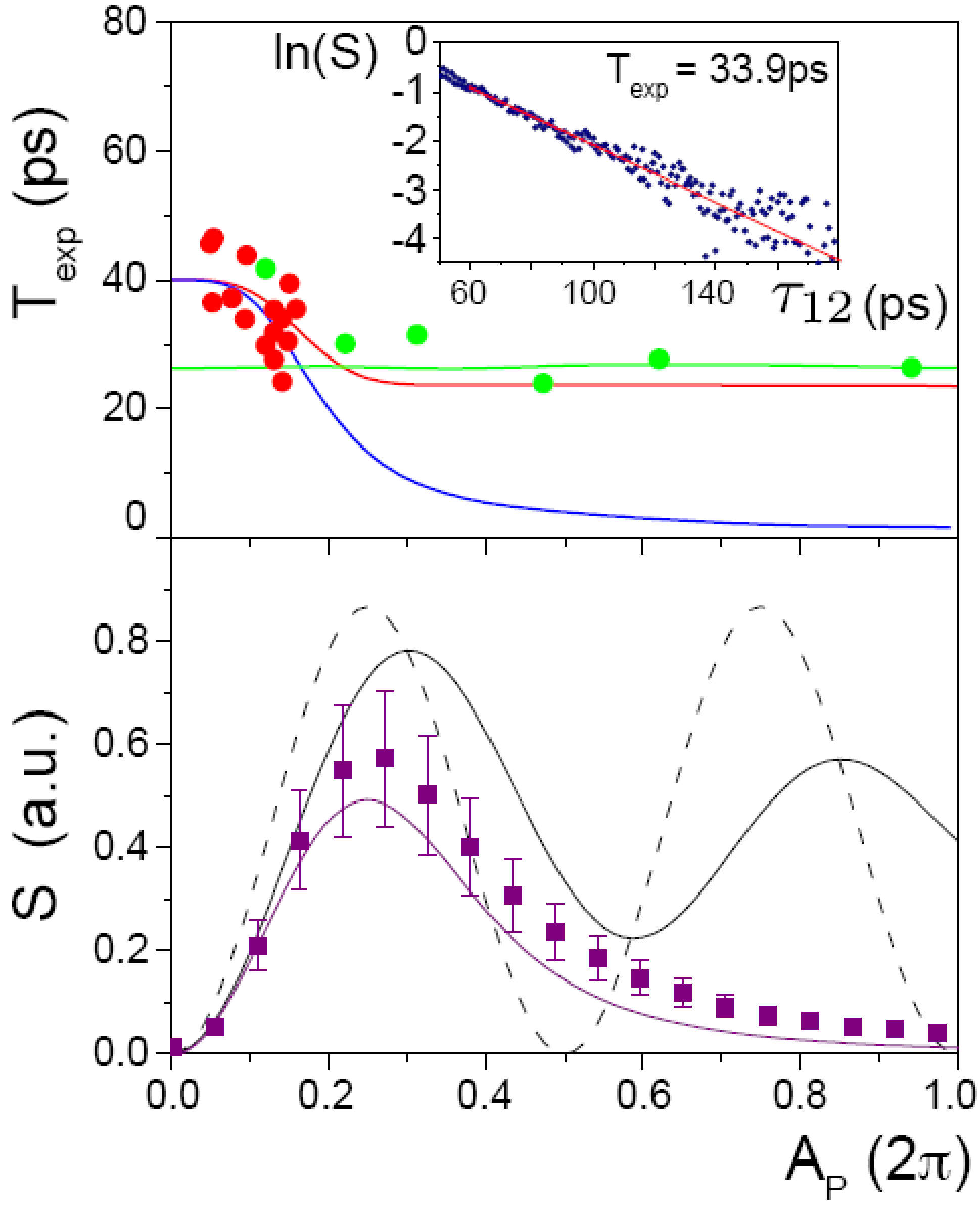

Fig. 3. (a) The echo dephasing lifetime as a function of pump pulse area $A_P$ for pump:rephasing pulse area ratio fixed at 1:2 and pulse duration of 5.9 ps (red points), and for rephasing pulse area fixed at 0.96 π for a pulse duration of 7.5 ps (green). The red and green lines show the corresponding results from the density matrix calculation described in the text. The blue line illustrates the effect on the red line of changing the radial profile of the laser beam to a top-hat

function. The inset shows a typical echo signal, detected via momentum selection (“ANGLE” in Fig 2a) with a constant background subtracted, as a function of the delay $\tau_{12}$ between pump and rephasing pulses. The data show a simple exponential decay with decay constant $T_{exp}$ (just as the time-resolved echoes of Fig 2e (b). Time-integrated photon echo signal *S* as a function of pump pulse area (in units of 2π) for a rephasing pulse area of 0.54 π and a pulse length of 6.79 ps. The dotted line is the uncorrected theoretical (Rabi) result. The black line shows the prediction including the non-uniform spatial profile of the laser beam, and the magenta line includes the effect of both photoionization and the profile. The experimental results for the same conditions are shown as points, with the ordinate normalized by the factor of 1.3 (determined from the global fit of many similar experiments with different pulse lengths and rephasing pulse areas - see supplementary material).

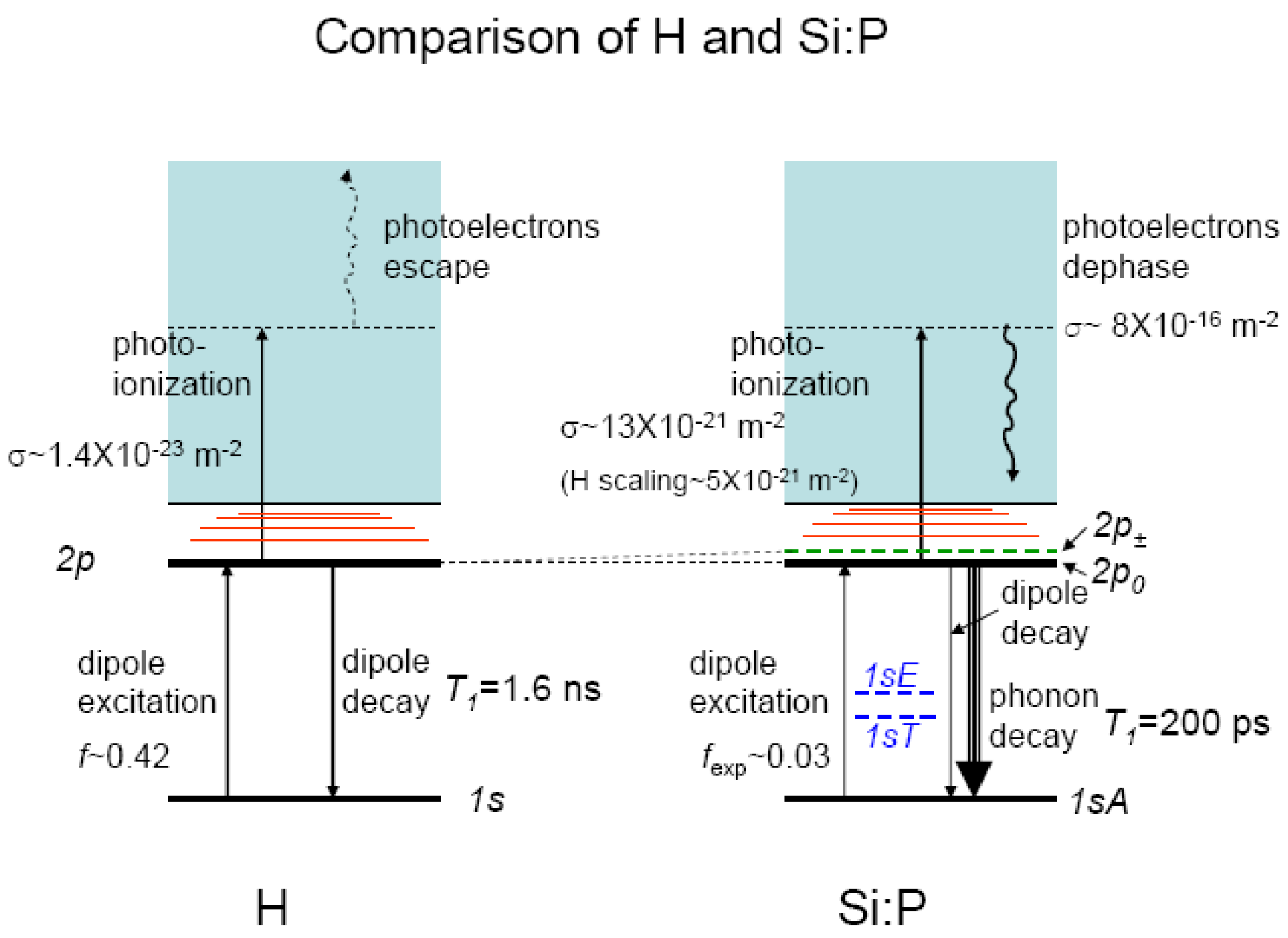

Fig 4. Comparison between H and Si:P, showing spectra and the principal excitation and decoherence mechanisms. The crystal environment leads to an asymmetric effective mass, so that the H 2p state splits into a $2p_0$ and $2p_\pm$ state (green dashed line). Additionally, the crystal field further splits the H-like levels into states of cubic symmetry – the splitting of the 1s level into its A, T and E components is shown, but the splitting in the excited states is too small to show. The main decay channel of the $2p_0$ state for Si:P, in contrast to H, is phonon decay: furthermore photoionization of this state is also relatively much stronger than in H, and since the photoelectrons are confined to the conduction band they can cause decoherence through phase changing collisions.

# Supplementary Material

The Dutch free-electron laser (FELIX) provides short coherent pulses of THz radiation. The laser pulses come in bursts called macropulses with duration of 5μs and repetition rate of 5 Hz. The macropulses contained trains of micropulses of duration ~ 10 ps and 25 MHz repetition rate. Figure 2a is a schematic diagram of our experiment. The incoming FELIX beam was directed onto a beam-splitter ($BS_1$). The transmitted beam which we call $\boldsymbol{k}_2$ travelled along a fixed optical path. The intensity of this beam could be controlled by a set of calibrated wire-grid attenuators (A). The reflected beam, $\boldsymbol{k}_1$ travelled down an optical path with controllable length ($R_1$) to provide a variable delay. It also passed through a variable attenuator consisting of a motorised polariser ($P_1$) and a fixed analyser ($P_2$). The $\boldsymbol{k}_1$ beam emerged from the analyser parallel to, but spatially separated from $\boldsymbol{k}_2$. All of these optical elements were contained in a closed metal box which was evacuated and purged several times with dry nitogen gas and then sealed. The spatially separated beams were reflected from a large off-axis parabolic mirror to a focus on the sample, which was mounted on the cold-finger of a continuous flow liquid helium cooled cryostat. The angular separation of the beams at the sample was ~5 degrees. The emerging pump and re-phasing beams and the echo beam were collected on a second off-axis parabolic mirror and collimated. A small mirror oriented at 45$^{\circ}$ and mounted on a translation stage was used to pick off the different beams for measurement of the echo angular position relative to the other beams. The light was focussed onto a helium-cooled Ge:Ga detector (crystal size 3x3 $mm^2$), by a third off-axis parabolic mirror.

To observe the pulse arrival times each of the beams ($\boldsymbol{k}_1$, $\boldsymbol{k}_2$, and $\boldsymbol{k}_E$) were made to interfere with a fourth, reference, pulse $\boldsymbol{k}^*$ derived from a small portion of $\boldsymbol{k}_2$. For this experiment we used the alternative detection setup shown in figure 2a TIMING; $\boldsymbol{k}^*$ was sent through a second optical delay line ($R_2$) and then recombined with the echo beam line through a second pellicle beam-splitter, and focused through a 0.4 mm diameter pinhole at the detector. The pinhole facilitated interference of the beams by restricting the number of fringes seen by the detector to one. The length of the optical delay was adjusted to give an appropriate interval between the excitation pulse $\boldsymbol{k}_1$ and the rephasing pulse $\boldsymbol{k}_2$. The optical delay line in the reference beam was scanned through its full range while pairwise combinations of the three main beams $\boldsymbol{k}_1$, $\boldsymbol{k}_2$ and the echo were blocked.

To control the intensity of the excitation pulse $\boldsymbol{k}_1$we used the motorised polariser shown in figure 2a. We calibrated the $\boldsymbol{k}_1$ intensity (in terms of pulse area) against the orientation angle of the polariser using a double Malus' Law (taking account of the polarisation orientation of FELIX with respect to the first polariser, and then the orientation of this polariser with respect to the analyser). We then recorded the intensity of the THz signal in the direction of the echo as a function of calibrated pulse area. We also recorded the intensity of the THz signal along the same direction when the temporal delay between $\boldsymbol{k}_1$ and $\boldsymbol{k}_2$ was reversed (i.e. we changed the length of the optical delay so that $\boldsymbol{k}_2$ arrives at the sample before $\boldsymbol{k}_1$). Since the echo is directional this provided us with a measure of the background signal from the scattered excitation pulse $\boldsymbol{k}_1$ and rephasing pulse $\boldsymbol{k}_2$. This was important because at high excitation powers the scattered signal began to exceed the echo signal.

For the echo lifetime experiments $\boldsymbol{k}^*$ was not used. Instead the total echo signal was measured as the pump-rephasing delay was varied, and the pump intensity controlled with the polarizer-analyser system. For some experiments both pump and rephasing beam were sent through the polarizer-analyser system, so that the ratio of their intensities was controlled by the beamsplitter (at 1:4). Similarly, the Rabi oscillation experiments were done by fixing the pump-rephasing pulse delay, and controlling the pump intensity with the polarizer-analyser, and the rephasing beam with the wire attenuator. The data presented here were obtained using a 200 μm thick CZ sample at a doping density of 1.5 X $10^{15}$ $cm^{-3}$, but we have also observed echos in both other CZ as well as FZ samples of broadly similar specification.

All our results can be described using a theoretical model which augments the standard echo theory with an extra source of decoherence due to photoionization. The basic information is in the two-level reduced density matrix ρ(*t*), which represents a very slight extension of the more usual Bloch equations. We have

$$\begin{pmatrix} \dot{\rho}_{12} \\ \dot{\rho}_{21} \\ \dot{\rho}_{11} \\ \dot{\rho}_{22} \end{pmatrix} = \begin{pmatrix} -\Gamma + i\Delta & 0 & i\Omega/2 & -i\Omega/2 \\ 0 & -\Gamma - i\Delta & -i\Omega/2 & i\Omega/2 \\ i\Omega/2 & -i\Omega/2 & 0 & \gamma \\ -i\Omega/2 & i\Omega/2 & 0 & -\gamma-\gamma' \end{pmatrix} \begin{pmatrix} \rho_{12} \\ \rho_{21} \\ \rho_{11} \\ \rho_{22} \end{pmatrix}$$

Here **ρ** is the 2×2 density matrix for the 2-level atom whose dynamics are being considered. The parameters have the following significance: γ is the rate at which the excited state population decays to the ground state — it is the reciprocal of $T_1$; *Γ* is the decoherence rate, i.e. $1/T_2$ . The instantaneous laser coupling is given by Ω/2 with

$$\Omega = eF(t)\mu_{12}/\hbar$$

where *F* is the (time-dependent) electric field envelope due to the laser and $\mu_{12}$ is the 1sA →$2p_0$ dipole moment so that Ω(t) is the instantaneous Rabi frequency. Of course, Ω is a function of time – it reflects the echo pulse sequence – though we have not explicitly indicated this, and $\Delta$ is the detuning of the laser from the atomic line centre. Finally $\gamma'$ is the rate of irreversible population loss from the upper level. For $\gamma'=0$ we have the usual Bloch equations, but we relate $\gamma'$ to the photoionization rate from level 2, so that

$$\gamma' = \sigma_{2p0} I(t) / \varepsilon$$

where $\sigma_{2p0}$ is the photoionization cross section from the excited state, $I(t)$ is the laser intensity at time $t$ and $\varepsilon$ is the photon energy (so $I(t)/\varepsilon$ is the photon flux). The photoelectrons thus produced also represent a source of decoherence which adds to the off-diagonal decay rate $\Gamma$, so that, for a photoelectron density $n_e$ given by $n_e = n_0 p_e$, where $n_0$ is the donor density, and $p_e=1-\rho_{11}-\rho_{22}$ is the electron ionization probability we have

$$\Gamma=\Gamma_0 + n_e v_e \sigma_e = \Gamma_0 + n_0 p_e v_e \sigma_e$$

with $\Gamma_0$ the intrinsic off-diagonal rate, $v_e$ the mean photoelectron velocity, and $\sigma_e$ the electron-donor scattering cross-section. This leads to an intensity-dependent echo decay rate. Finally, to compare with experiment we must average over the spatial distribution of the FELIX beam. This model can then be used as the basis of a fit to the experimental profiles as a function of pulse area. Figure S1 shows the global comparison between the theoretical and experimental results.

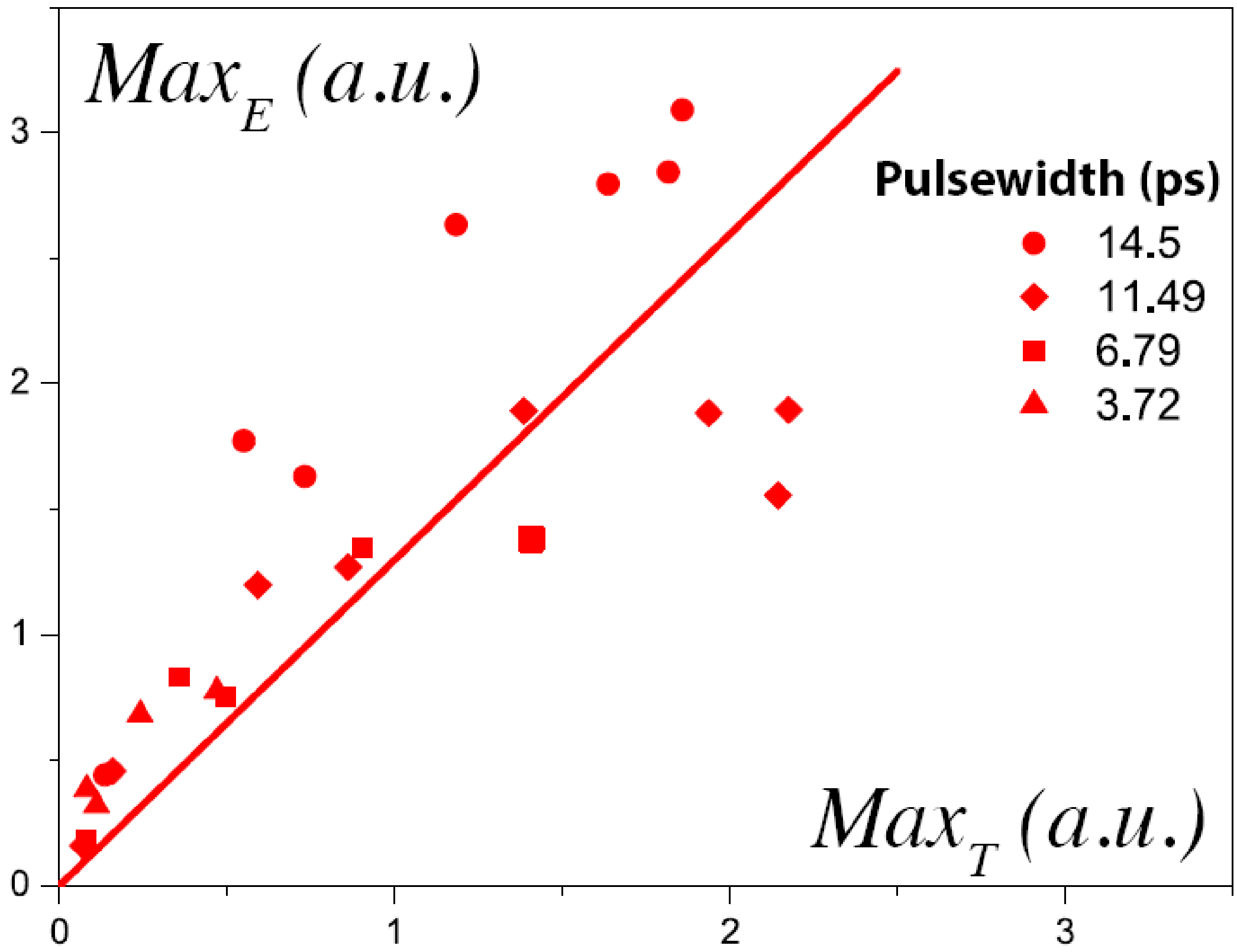
Max_E (a.u.)
Max_T (a.u.)
Pulsewidth (ps)
14.5
11.49
6.79
3.72
0
1
2
3
0
1
2
3

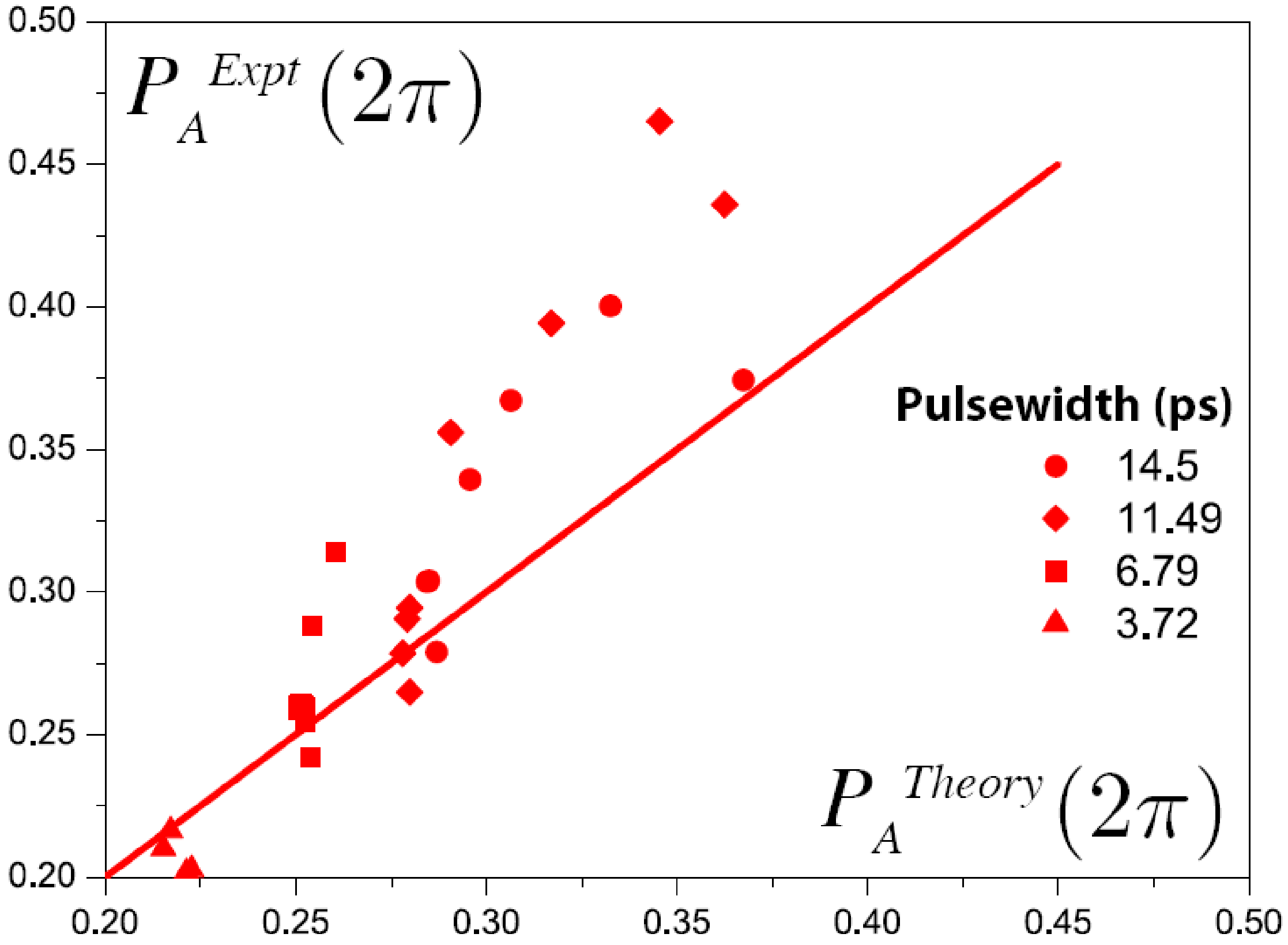

Fig S1  Comparison between theory and experiment. The upper panel shows the value of the experimental maximum of the echo signal intensity (such as that in fig 3b) (ordinate) as a function of the corresponding theoretical value (abscissa).  The best fit line through the origin has a gradient of 1.3±0.06, which gives the best calibration of the experimental echo magnitude in terms of the theoretical value as calculated above.

The lower panel compares the experimental value of the pump pulse area (in units of 2π) for which the peak echo signal occurs (again taken from profiles such as that shown in fig 3b) (ordinate) with the theoretical prediction (abscissa) The line through the origin with gradient 1 is also shown. (It is not a fit.)